# The electronic structure of polyaniline and doped phases studied by soft X-ray absorption and emission spectroscopies

M. Magnuson, J.-H. Guo, S. M. Butorin, A. Agui, C. Såthe and J. Nordgren

*Department of Physics, Uppsala University, P. O. Box 530, S-751 21*

*Uppsala, Sweden*

A. P. Monkman

*Department of Physics, University of Durham, South Road, Durham DH1 2UE, England*

**Abstract**

The electronic structure of the conjugated polymer, polyaniline, has been studied by resonant and nonresonant X-ray emission spectroscopy using synchrotron radiation for the excitation. The measurements were made on polyaniline and a few doped (protonated) phases for both the carbon and nitrogen contents. The resonant X-ray emission spectra show depletion of the $\pi$ electron bands due to the selective excitation which enhances the effect of symmetry selection rules. The valence band structures in the X-ray emission spectra attributed to the $\pi$ bands show unambiguous changes of the electronic structure upon protonation. By comparing to X-ray absorption measurements, the chemical bonding and electronic configuration is characterized.

## 1 Introduction

Conjugated polymers have the electronic structure of quasi-one-dimensional electron phonon-coupled semiconductors and have been intensively studied since the conductivity induced by doping was discovered [1,2,3]. By doping the conjugated polymer, a charge transfer is introduced through either oxidation or reduction (n or p-type of doping). In the case of polyaniline (PANi), doping is achieved by protonation of backbone nitrogen sites. Thus the total electron number does not change but vacancies (on two sites) are created.

Most of the interesting chemistry and physics of conducting polymers are associated with the electronic structure at the $\pi$-levels at the valence and conduction band edges. Spectroscopic techniques, such as photoelectron spectroscopy in the X-ray and ultraviolet wavelengths regimes have been employed for studies of the uppermost valence band $\pi$-levels [4,5]. Among the various available conducting polymers, PANi is of particular interest since it has a rather high electrical conductivity in the doped phases, and is relatively easy to synthesize.

Figure 1 shows the basic chemical structure of one repeat unit of PANi [6,7,8,9,10]. In each repeat unit there are three benzene rings (denoted 1-3 in Fig. 1) separated by amine ($-NH$) groups and one quinoid ring (4) surrounded by imine ($-N=$) groups. For the quinoid ring which forms double bonds with the nitrogens, there are two pairs of carbon atoms in the ring and four $\pi$-electrons. The macroscopic structure of the polymer is made up of many long chains of repeat units forming a complicated network. PANi exists in three different forms depending on the oxidation state; leucoemeraldine (y=1), emeraldine (y=0.5) and pernigraniline (y=0). Each form can exist in a base form and in various protonated $H^+$ salt forms. Only the protonated salt form of emeraldine is conductive. Upon protonation, the quinoid-imine group transforms into a semiquinone radical cation state, i.e., emeraldine salt.

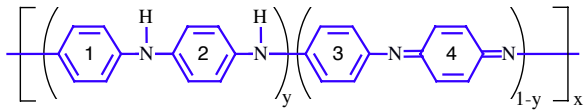

**Figure 1:** The basic geometrical structure of polyaniline. In the emeraldine form y=0.5 [6].

In the present work, we use soft X-ray emission (SXE) spectroscopy to investigate the electronic structure of neutral and different doped phases of the protonated emeraldine form (y=0.5) of PANi. The PANi samples were doped via protonation with camphorsulphonic acid (CSA) and acrylamedo-2-proparesulphic acid (AMPSA). The protonation adds $H^+$ so that holes are added to the imine groups around the quinoid ring which can be looked upon as a charge delocalization of holes in the valence band. When the polymer is doped, the double-bonds reorganize and there are thus three pairs of carbon atoms in each ring with a total of six π-electrons. In the undoped (base) form of PANi, the amine groups contain non-bonding (lone-pair) nitrogen orbitals while the imine groups contain a mixture of lone-pair and π-conjugated orbitals. In the doped phases the quinoid-imine double bonds become semiquinone, the chemical bonds rearrange and the torsion angles between the benzene rings become nearly planar. Upon doping with either CSA or AMPSA, which is performed in solution to achieve homogeneous doping, the acid hydrogen ion, bonds to the imine sites of the polymer chain. On casting of films from such solutions, this interaction induces crystallinity in the samples [11] and gives rise to metallic electrical properties [12]. The SXE technique provides a means of extracting electronic structure information in terms of local contributions to the Bloch or molecular valence orbitals, since the core-hole decay can be described according to the dipole selection rules [13,14]. However, the relatively low fluorescence yield and instrumental efficiencies associated with SXE in the ultra-soft X ray regime places considerable demands on the experimental measurements in this spectral region [15,16]. In the following section, we describe the details of the experiment (section 2), followed by a presentation of the results and discussion (section 3).

## 2 Experimental Details

The experiment was carried out at beamline 7.0 at the Advanced Light Source (ALS) at the Lawrence Berkeley National Laboratory (LBNL). This undulator beamline includes a spherical-grating monochromator [17] and provides linearly polarized synchrotron radiation of high resolution and high brightness. Near-edge X-ray absorption fine structure (NEXAFS) [5] spectra were recorded by measuring the total electron yield with 0.25 eV and 0.40 eV resolution of the beamline monochromator for both the carbon and nitrogen edges, respectively. The NEXAFS spectra were normalized to the incident photon current using a gold mesh inserted in the excitation beam.

The SXE spectra were recorded using a high-resolution grazing-incidence X-ray fluorescence spectrometer [18]. During the SXE measurements, the resolution of the beamline monochromator was the same as in the NEXAFS measurements. The X-ray fluorescence spectrometer had a resolution of 0.30 eV and 0.65 eV, for the carbon and nitrogen measurements, respectively. The energy scales of the NEXAFS spectra were calibrated using the elastic peaks in the SXE spectra. The incidence angle of the photons was 20 degrees with respect to the sample surface which was vertically oriented. During the data collection, the samples were scanned (moved every 30 seconds) to avoid the effects from photon-induced decomposition of the polymers.

The PANi samples were synthesized chemically using the standard Durham route [19]. The undoped (base) emeraldine films were prepared from NMP solutions, n-methyl-2-pyrolidinone [20]. The polymer was disolved in appropriate solvents with CSA or AMPSA [12] and films were cast onto silicon wafers. After drying the films were peeled off. For the measurements, the films were mounted with double adhesive carbon tape and inserted into a vacuum chamber. The base pressure in the experimental chamber was $4\times10^{-9}$ Torr during the measurements.

# 3 Results and discussion

## 3.1 $C_{1s}$ X-ray absorption spectra

Figure 2 shows a set of NEXAFS spectra measured at the C 1$s$ threshold of PANi and the doped phases. An intense peak at about 285.0 eV (denoted 1) corresponds to the lowest unoccupied $\pi^*$ molecular orbitals. At higher photon energies, a shape resonance (4) with $\sigma^*$ character and superimposed multielectron excitations is observed. For a qualitative understanding of the different transitions involved in the polymer, a C 1$s$ NEXAFS spectrum of condensed aniline (aminobenzene) is also shown at the bottom for comparison [21]. The C 1$s$ NEXAFS spectrum of condensed aniline is dominated by a $\pi^*$ resonance which is represented by two peaks (1 and 2) due to a chemical shift as a result of the carbon atoms being either bonded or unbonded to $NH_2$. The first $\pi^*$ peak (1) arises from three levels corresponding to the unbonded carbon atoms located at about 285.4 eV while the second peak (2) contains only one level located at 286.8 eV corresponding to the carbon atom with the nitrogen as nearest neighbour [21]. The $NH_2$ group causes the separation of the orbitals of the carbon atoms, and the degeneracy of the $\pi^*$ molecular orbitals of the benzene rings decreases. The third feature (3) can probably also be related to ring carbon excitations of $\pi$ character [22,23].

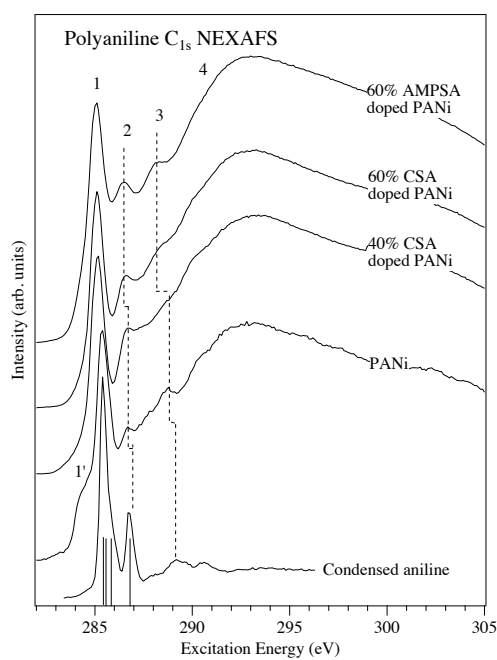

**Figure 2:** A series of NEXAFS spectra recorded at the carbon edge for neutral PANi base, different doped phases of PANi and for condensed aniline.

In the C 1$s$ NEXAFS spectrum of the undoped PANi polymer, similar peaks are observed as in the condensed aniline, although the structures are broader due to a larger number of transitions. The alternating carbon and nitrogen atoms in each unit cell gives rise to a chemical shift and the second $\pi^*$ peak (2) at about 286.5 eV. In the pre-threshold region of the undoped PANi, a shoulder (1') is observed at about 284.5 eV which can be attributed to those carbon atoms bonded to nitrogen via double bonds in the quinoid ring. From optical absorption studies [24], it is known that the quinoid-imine parts absorb at 2 eV compared to the 3.6 eV $\pi \rightarrow \pi^*$ transition of the benzenoid-amine parts, a shift of 1.6 eV consistent with a low-energy excitonic shift in the present NEXAFS data. The doped phases show similar peak structures as the undoped PANi, but the peaks are shifted towards lower energies. Peak (1') dissappears in the doped phases indicating a breakup of the quinoid-imine double bonds at the nitrogen sites upon protonation.

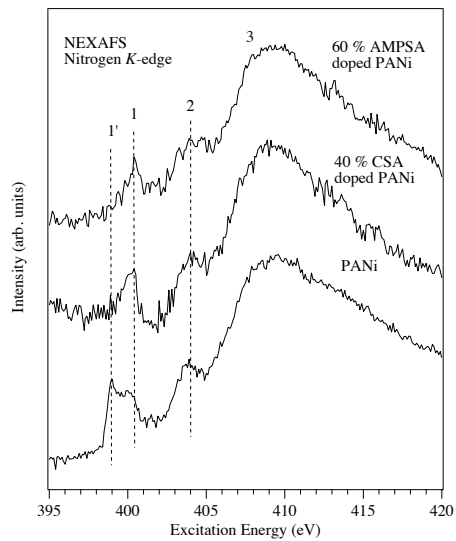

**Figure 3:** A series of NEXAFS spectra recorded at the nitrogen edge for neutral PANi base and two doped phases of PANi.

## 3.2 $N_{1s}$ X-ray absorption spectra

Figure 3 shows NEXAFS spectra of PANi measured at the N 1s threshold. In the undoped PANi spectrum (bottom), the first peak is split up into two peaks (1 and 1') of nearly equal intensity at about 399.0 and 400.5 eV photon energy. These superimposed structures have been ascribed to $\pi^*$ imine (1') and amine (1) groups, respectively [6]. The splitting of the $\pi^*$ peak arises from the mesomeric conjugation effect. By measuring the relative $\pi^*$-electron density in this way, it is thus possible to distinguish amine nitrogen from nitrogen with a mixture of imine and amine bonding. The undoped PANi sample thus contain a rather high degree of imine nitrogens.

In the NEXAFS spectrum of the 40 % CSA doped sample, the low energy part of peak (1') assigned to the $\pi^*$ imine group almost disappear. Only a small low-energy shoulder on the 400.5-eV structure (1), can possibly be related to the imine nitrogens. In the spectrum of the 60 % AMPSA PANi sample, this peak (1) is sharper, indicating that it is only due to amine nitrogens. The second and third peaks (2 and 3) visible in all three spectra can be related to delocalized $\sigma^*$ resonances [25].

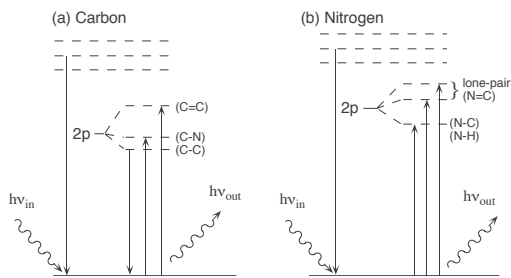

**Figure 4:** Schematic energy level diagram at the (a) carbon and (b) nitrogen *K*-edges.

## 3.3 Resonant and nonresonant X-ray emission spectra at the carbon *K*-edge

Fig. 4 shows a schematic energy level diagram of the transitions involved in the X-ray transitions. The X-ray emission process of PANi is governed by the 1s → 2p dipole transitions between the localized core level and the valence band. Figures 5 and 6 show non-resonant and resonant C *K* X-ray emission spectra of undoped PANi and the doped phases measured with the excitation energies of 302.2 eV and 285.0 eV. By comparing the observed emission spectra with molecular orbital calculations [21], and bandstructure calculations [26], three major bands (labeled A-C) can be identified in the spectra. Peak A corresponds to $\pi$-like states, peak B is due to both $\pi$ and $\sigma$ contributions, with $\sigma$ dominating, while peak C is due only to $\sigma$ electronic states. For the condensed aniline (bottom), the features are somewhat more prominent since there are less transitions than in the case of a polymer.

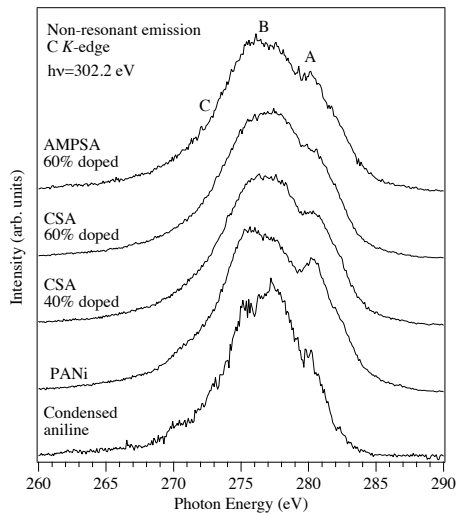

**Figure 5:** Non-resonant X-ray emission spectra measured at the carbon $K$-edge.

It should be mentioned that the three bands A-C observed in both the nonresonant and resonant carbon X-ray emission spectra, correspond well in energy with the three bands observed in photoemission spectra of PANi reported by H. Sakamoto et al. [7]. In valence band photoemission spectroscopy, the same final states are obtained as in the non-resonant carbon $1s$ X-ray emission spectra.

Comparing the spectra of the neutral and doped samples of PANi in Fig. 5, a significant change of the shape of band A can be identified depending on the doping level. A comparison of the relative intensities of the $\pi$ and $\sigma$ bands show that the $\pi$-band is more intense in the undoped sample (PANi base), than in the doped samples. In the nonresonant spectra of the undoped polymer, the A/B peak-ratio is 0.8, whereas for the 40% CSA doped sample, this ratio is reduced to 0.7. The effect is about the same for the 60% doped CSA and AMPSA samples where the A/B peak-ratio is also about 0.7. Thus the same effect is achieved with AMPSA and CSA at the same doping level. The difference between the doping levels can be interpreted as an increase of the number of holes in the valence band upon protonation on the nitrogen sites (p-doping).

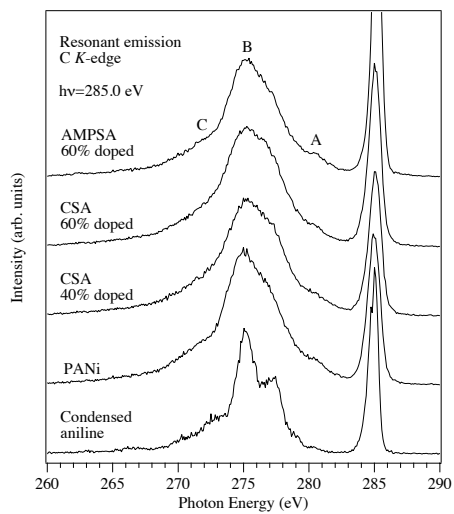

**Figure 6:** Resonant X-ray emission spectra at the carbon $K$-edge.

The resonant C $K$ spectra in Fig. 6 basically consist of two main features: the inelastic scattering contribution and the elastic peak with a position equal to the excitation energy. The $\pi$-bands (peak A and part of peak B) are dramatically depleted in the resonant case as a result of the symmetry selection rules [15,27]. Only a very small structure of peak A remains which may be related to a carbon lone-pair orbital. A similar effect is observed for the condensed aniline where also the $\pi$-band part of peak B is clearly depleted which has been well reproduced by molecular orbital calculations [21]. The elastic peaks in the doped samples are somewhat higher than in the undoped case. The strength of the elastic peak reflects to some extent the degree of localization of the $\pi^*$ orbitals.

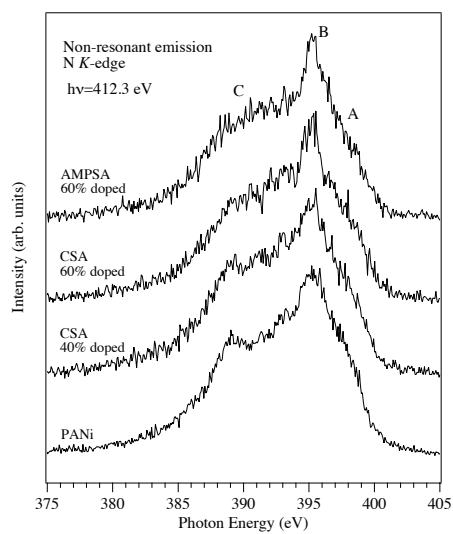

**Figure 7:** Non-resonant X-ray emission spectra at the nitrogen *K*-edge.

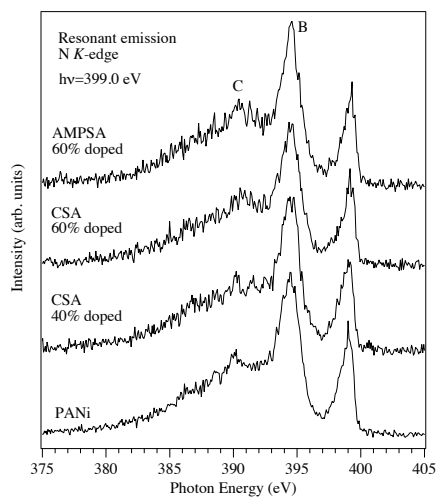

**Figure 8:** Resonant X-ray emission spectra at the nitrogen *K*-edge.

Comparing the PANi emission spectra to those of poly-pyridine (PPy) [16], the peak structures are broader since the geometrical structure of the unit cell of the PANi polymer is more complex. In addition, the C 1*s* core-level shifts introduced by the nitrogen atoms are larger in PPy since the nitrogen atoms are located inside the benzene rings [16].

### 3.4 Resonant and nonresonant X-ray emission spectra at the nitrogen *K*-edge

Figures 7 and 8 show non-resonant and resonant N *K* emission spectra of PANi and the doped samples, excited with 412.3 eV and 399.0 eV photon energies, respectively. The N 1*s* X-ray transitions and core level splittings are schematically shown in Fig. 4. The largest difference of the nonresonant and resonant nitrogen spectra in comparison to the carbon spectra is the relatively strong intensity of band B. In the resonant nitrogen spectra, the elastic peak is identified at 399.0 eV, which is the same energy as the incoming photon energy. The elastic peak is relatively lower at the N *K*-edge than at the C *K*-edge. In Fig. 8, the A-peak is almost entirely depleted due to the symmetry selection rules. For the A-peak in Fig. 7, the same effect is observed as in the carbon spectra; it is somewhat smaller in the doped samples since there are less occupied states at the uppermost π-levels than in the undoped sample. In the case of molecular aniline [28], and PPy [16], peak B has been interpreted as corresponding to the nitrogen 2*p* non-bonding lone-pair orbital. Comparing the nonresonant and the resonant nitrogen *K*-emission spectra, the low-energy part of the B-structure remains while the high-energy part A is depleted in the resonant spectra. This is probably due to the fact that there is a high degree of mixing of ring π and lone-pair orbitals in the B-band. In the case of PPy the lone-pair nitrogen orbitals are much less mixed into the π-electron orbitals [16].

From Figs. 7 and 8, it is also clear that protonation enhances the intensity of band B and gives a relative decrease of the A-band. As mentioned previously, the decrease of the A-band upon doping is due to a depopulation of the uppermost π-levels. It is interesting to note that the B-band is somewhat enhanced upon protonation. The major part of the B-band probably originates from the amine nitrogen lone-pair orbitals which are not affected by protonation. However, since the B-peak probably also contain some fraction of π-orbitals, these are depleted upon protonation which enhances the lone-pair effect. The A-peak probably originates to a higher degree from the imine groups which obviously are reduced on protonation.

The reason why we do not observe a depletion of peak B going from the non-resonant to the resonant spectra in the nitrogen case is due to the fact that the strong B-band in those spectra is dominated by the lone-pair orbital localized to nitrogen, which has σ symmetry [16]. It is also interesting to note that the lone-pair contribution is not depleted when the sample is doped. Thus, upon protonation of the nitrogen sites, this results in a pure charge-transfer from the nitrogen to the protons.

# 4  Summary

We present resonant and non-resonant X-ray emission measurements of the emeraldine form of polyaniline and doped (protonated) phases. The most prominent difference between the resonant and non-resonant spectra is the depletion of the π orbitals due to the energy selectivity which enhances the effect of the symmetry selection rules. A detailed comparison of the X-ray emission spectra from different samples shows that there are differences in the electronic structure depending on the doping level. The protonated phases of polyaniline were found to have less occupied states at the uppermost π-levels as a result of the charge transfer from the protonated nitrogen sites. A comparison of the carbon and nitrogen spectra show that the orbitals have different relative intensities. In particular, the non-bonding (lone-pair) levels mixed into the π-levels are emphasized in the nitrogen spectra.

# 5  Acknowledgments

This work was supported by the Swedish Natural Science Research Council (NFR), the Swedish Research Council for Engineering Sciences (TFR), the Göran Gustavsson Foundation for Research in Natural Sciences and Medicine and the Swedish Institute (SI). The ALS, Lawrence Berkeley National Laboratory was supported by the U. S. Department of Energy, under contract No. DE-AC03-76SF00098.

# References

[1]   C. K. Chiang, C. R. Jr. Fincher, Y. W. Park, A. J. Heeger, H. Shirakawa, E. J. Louis, S. C. Gau and A. G. MacDiarmid; Phys. Rev. Lett. **39**, 1098 (1977).

[2]   J. H. Burroughes, D. D. C. Bradley, A. R. Brown, R. N. Marks, K. Mackay, R. H. Friend, P. L. Burns, and A. P. Holmes; Nature **347**, 539 (1990).

[3]   F. Garnier, G. Horowitz, X. Peng and D. Fichou; Adv. Mater. **2**, 592 (1990).

[4]   A. B. Holmes, D. D. C. Bradley, A. R. Brown, P. L. Burn, J. H. Burroughes, R. H. Friend, N. C. Greenham, R. W. Gymer, D. A. Halliday, R. W. Jackson, A. Kraft, J. H. F. Martens, K. Pichler and I. D. W. Samuel; Synth. Met. **54**, 401 (1993).

[5]   J. Stöhr, *NEXAFS spectroscopy* (Springer-Verlag, Berlin 1992).

[6]   C. Hennig, K. H. Hallmeier, R. Szargan; Synth. Met. **92**, 161 (1998).
[7]   H. Sakamoto, M. Itow, N. Kachi, T. Kawahara, K. Mizoguchi, H. Ishii, T. Tiyahara, K. Yoshioka, S. Masubuchi, S. Kazama, T. Matsushita, A. Sekiyama and S. Suga; J. Electr. Spec. **92**, 159 (1998).
[8]   A. P. Monkman, G. C. Stevens and D. Bloor; J. Phys. D: Appl. Phys. **24**, 738 (1991).
[9]   A. Kenwright, W. J. Feast, P. Adams, A. Milton, A. P. Monkman and B. Say; Polymer **33**, 4292 (1992).
[10]  S. J. Pomfret, P. N. Adams, N. P. Comfort and A. P. Monkman; Advanced materials **10**, 1351 (1998).
[11]  L. Abell, P. N. Adams and A. P. Monkman; Polymer Comm. **37**, 5927 (1996).
[12]  P. N. Adams, P. Devasagayam, S. J. Pomfret, L. Abell and A. P. Monkman; J. Phys. Condens. Matter. **10**, 8293 (1998).
[13]  H. M. O'Bryan and H. W. B. Skinner; Phys. Rev. **45**, 370 (1934).
[14]  H. Jones and N. F. Mott, and H. W. B. Skinner; Phys. Rev. **45**, 379 (1934).
[15]  J.-H. Guo, M. Magnuson, C. Såthe, J. Nordgren, L. Yang, Y. Luo, H. Ågren, K. Z. Xing, N. Johansson, W. R. Salaneck and W. J. Feast; J. Chem. Phys., **108**, 5990 (1998).
[16]  M. Magnuson, L. Yang, J.-H. Guo, C. Såthe, A. Agui, J. Nordgren, Y. Luo, H. Ågren, N. Johansson, W. R. Salaneck, L. E. Horsburgh and A. P. Monkman; Chem. Phys., **237**, 295 (1998).
[17]  T. Warwick, P. Heimann, D. Mossessian, W. McKinney, and H. Padmore; Rev. Sci. Instr. **66**, 2037 (1995).
[18]  J. Nordgren and R. Nyholm, Nucl. Instr. Methods A 246, 242 (1986); J. Nordgren, G. Bray, S. Cramm, R. Nyholm, J.-E. Rubensson, and N. Wassdahl; Rev. Sci. Instr. **60**, 1690 (1989).
[19]  P. Adams, P. J. Laughlin, A. P. Monkman and A. M. Kenwright; Polymer **37**, 3411 (1996).
[20]  A. P. Monkman and P. Adams, Synth. Met. **40**, 87 (1991).
[21]  Y. Luo, H. Ågren, J.-H. Guo, P. Skytt, N. Wassdahl and J. Nordgren; Phys. Rev. A **52**, 3730 (1995).
[22]  C. C. Turci, S. G. Urguhart and A. P. Hitchcock; Can. J. Chem. **74**, 851 (1996).
[23]  O. Plashkevych, Li Yang, O. Vahtras, H. Ågren and L. M. Pettersson; Chem. Phys. **222**, 125 (1997).
[24]  A. P. Monkman in *Conjugated Polymeric Materials*, ed. J. L. Bredas and R. R. Chance, NATO ASI Series E: Applied Science Vol. 182 p. 273, Kluwer Academic Publishers 1990.
[25]  C. Hennig, K. H. Hallmeier, A. Bach, S. Bender, R. Franke, J. Hormes and R. Szargan; Spectrochimica Acta Part A **52**, 1079 (1996).
[26]  J. L. bredas, B. Themans, J. M. Andre; J. Chem. Phys. **78**, 6137 (1983); J. L. Bredas, B. Themans, J. G. Fripiat, J. M. Andre; Phys. Rev. B **29**, 6761 (1984); S. Stafstrom, J. L. Bredas, A. J. Epstein, H. S. Woo, D. B. Tanner, W. S. Huang, A. G. MacDiarmid; Phys. Rev. Lett. **59**, 1464 (1987).
[27]  P. Skytt, J.-H. Guo, N. Wassdahl, J. Nordgren, Y. Luo, and H. Ågren; Phys. Rev. A **52**, 3572 (1995).
[28]  J. Nordgren, L. Selander, L. Pettersson, R. Brammer, M. Backström and C. Nordling; Chem. Phys. **84**, 333 (1984).